\documentclass[prl]{revtex4}%
\usepackage{amsfonts}
\usepackage{amsmath}
\usepackage{amssymb}
\usepackage{graphicx}%
\begin{document}
\preprint{}
\title[High Magnetic Field Studies...]{High Magnetic Field Studies of the Hidden Order Transition in URu$_{2}$%
Si$_{2}$}
\author{M. Jaime$^{1}$, K.H. Kim$^{1}$, G. Jorge$^{1,2}$, S. McCall$^{3}$, J.A.
Mydosh$^{4,5}$}
\affiliation{$^{1}$MST-NHMFL, Los Alamos National Laboratory, Los Alamos, NM 87545.\\ 
$^{2}$Departamento de F\'{\i}sica, Universidad de Buenos Aires, Bs.As., Argentina. \\
$^{3}$National High Magnetic Field Laboratory, Tallahassee, FL 32310. \\
$^{4}$Kamerlingh Onnes Laboratory, Leiden University, Leiden, The Netherlands. \\
$^{5}$Max-Planck-Institute f\"{u}r Chemische Physik fester Stoffe, Dresden, Germany.}


\begin{abstract}
We studied in detail the low temperature/high magnetic field phases of
URu$_{2}$Si$_{2}$ single crystals with specific heat, magnetocaloric effect,
and magnetoresistance in magnetic fields up to 45 T. Data obtained down to 0.5
K, and extrapolated to T = 0, show a suppression of the hidden order phase at
H$_{o}$(0) = 35.9 $\pm$ 0.35 T and the appearance of a new phase for magnetic
fields in excess of H$_{1}$(0) = 36.1 $\pm$ 0.35 T observed \textit{only} at
temperatures lower than 6 K. In turn, complete suppression of this high field
state is attained at a critical magnetic field H$_{2}$(0) = 39.7 $\pm$ 0.35 T. 
No phase transitions are observed above 40 T. We discuss our results in the context 
of itinerant vs. localized \textit{f}-electron behavior and consider the 
implications for the hidden order phase.

\end{abstract}
\volumeyear{year}
\volumenumber{number}
\issuenumber{number}
\eid{identifier}
\date{17 September 2002}
\received[Received]{10 May 2002}

\revised[Revised]{...}

\accepted[Accepted]{...}

\published[Published]{...}

\startpage{101}
\endpage{102}
\maketitle

During the past few years there has been a true renaissance of interest in the
unusual phase transition \cite{amitsuka00} that occurs in the superconducting
heavy fermion system URu$_{2}$Si$_{2}$ at T$_{o}$ $\approx$ 17 K, where all of
the thermodynamic and transport properties exhibit a mean-field-like anomaly
\cite{palstra85,brohlom87,maclaughlin88}. These early experimental results led
to the conclusion that a magnetic phase transition, possibly of a spin density
wave-type, took place. However, when probed with microscopic measurements,
\textit{e.g.}, neutron diffraction and muon spin rotation ($\mu$SR), only a
very tiny magnetic moment of $\approx$ 0.02 $\mu_{B}$/U was found. Such a
small moment could \emph{not} account for the large changes in behavior at the
phase transition and it gradually became apparent that an unconventional type
of magnetic order is at play. Indeed, recent neutron diffraction
\cite{amitsuka99}, nuclear magnetic resonance \cite{matsuda01}, and $\mu$SR
\cite{amitsukatbp} under pressure show the apparent tiny homogeneous moment to
be due to a metallurgical minority phase of large moments ($\approx$ 0.3
$\mu_{B}$), that coexists with a majority (bulk) phase that has no magnetic
moments. After more than 15 years of investigation the nature of the bulk
phase transition remains unidentified, and the term \emph{hidden order} (HO)
has recently been coined to describe this phenomenon\cite{shah00}. Besides
pressure, yet another external parameter is known to affect the ordered
state,\emph{\ i.e.} external magnetic fields. Pulsed-field measurements of
magnetization \cite{deboer86,sugiyama90,sakakibara93}, resistivity
\cite{devisser86,bakker93}, Hall effect \cite{schoenes87,bakker93}, and
ultrasonic velocity changes (elastic moduli, $c_{ij}$) \cite{wolf01} exhibit a
three-step transition between 35 and 40 T for which a satisfactory explanation
is still pending.

In this Letter we present the first measurements of the temperature -- field
dependences of the specific heat, magnetocaloric effect and resistivity from
0.5 K to 20 K with fields up to 45 T to attempt to address the outstanding
questions discussed above. The measurements indicate four regimes of anomalous
behavior as URu$_{2}$Si$_{2}$ emerges from its hidden ordered state: (I) The
continuous phase transition becomes sharper and symmetric in temperature as
magnetic field is increased above 32 T. (II) There is no phase transition to
be seen down to 0.5 K at $\sim$36 T. (III) Between 36 and 39 T a first
order-like transition reappears, and (IV) Above 40 T a Schottky-like maximum
develops without any sign of a phase transition. These characteristics, never 
observed before, can
then establish the basic ingredients of the HO state and form a critical test
for the correct theoretical description. We consider two different scenarios
to explain these behaviors: \textbf{A}) Zeeman splitting of itinerant
\textit{f}-electron bands suppress the HO phase in region (I), which reappears
as a single-spin ordered phase or an \textit{orbital-flop} phase (the
orbital-current equivalent to an AF spin-flop phase) in region (III). This
scenario is related to the exotic density wave mechanism recently proposed by
Chandra et al. \cite{chandra02} although these authors have not yet considered
possible high magnetic field transitions. \textbf{B}) Crossing of
\textit{f}-electron crystal electric field (CEF) levels induce a quadrupolar
ordered phase at high fields, region (III). Here we have the localized model
of Santini \cite{santini98} which ignores correlation effects between
\textit{f}-electrons. Our data also suggest that quantum bi-criticality may
lie in the middle of the field region (II), \emph{i.e.} where T$_{o}$%
(H$\simeq$36T)$=$ 0.

Two different samples were measured, sample $\sharp$1 was used for specific
heat vs. T and magnetoresistance measurements, sample $\sharp$2 for
magnetocaloric effects. Single crystals of sample $\sharp$1 were fabricated by
tri-arc melting (Czochralski method) stoichiometric amounts of U, Ru and Si.
After the growth process the compound was annealed at 950 $^{o}$C for one
week. The crystal was characterized by Debye-Scherrer and Laue X-ray
diffraction, and electron probe microanalysis. These results showed the
crystal to be of excellent quality: on stoichiometry and no second structural
phases. Measurements of the specific heat and magnetoresistance, that show HO
transition at T$_{o}$=17.1 K, were carried out on oriented plate-like and
bar-like samples, formed by spark erosion so that the external field is always
parallel to the tetragonal $c$-axis. Single crystals of sample $\sharp$2 ,
T$_{o}$ = 17.4 K, were grown by arc-melting followed by vertical float-zone
refining as described elsewhere \cite{kwok90}.

The specific heat of sample $\sharp$1 (see Figure 1) was measured on a
bar-shaped 9 mg piece with the $c$-axis along the bar principal axis. We 
used a standard thermal
relaxation method, with both \emph{small} and \emph{large} delta T
\cite{bachmann72}, to determine the specific heat as a function of the
temperature at constant magnetic fields up to 45 T. The temperature was
measured with a Cernox bare chip resistance thermometer (Lakeshore Inc.)
calibrated as described before \cite{jaime00}. Measurements were performed at
the National High Magnetic Field Laboratory (NHMFL), Tallahassee, in both a
water cooled resistive magnet operating to 32 T, and a hybrid magnet
consisting of a resistive coil inside a superconducting coil operating to a
total field of 45 T. From the total specific heat measured (C$_{tot}$) we
subtracted the contribution from phonons (C$_{ph}$) measured in a sample of
ThRu$_{2}$Si$_{2}$ \cite{amitsuka94}. Figure 1 (a) displays C$_{m}%
$/T=(C$_{tot}$-C$_{ph}$)/T vs. temperature for magnetic fields up to 33.5 T.
Our data at low fields is in excellent agreement with previous measurements
\cite{vandijk97}. We observe that the anomaly associated to the HO phase in
URu$_{2}$Si$_{2}$ is shifted to lower temperatures by the external magnetic
field, becoming sharper and more symmetric, without changing the amount of
entropy recovered at the transition, which remains close to 0.15 R (where R is
the Rydberg gas constant). The sharpening of the anomaly indicates a gradual
switch from continuous (second order) to discontinuous (first order) in
temperature, however, we do not observe the hysteresis expected for such a
transition. Figure 1 b) displays C$_{m}$/T measured at 36 T and 38 T. We see
here the complete suppression of the peak in C$_{m}$/T heat associated with
the HO phase. Indeed, the data at H = 36 T show only a small step feature
resembling that of CeRu$_{2}$Si$_{2}$ near the metamagnetic transition at
H$_{m}$ = 7.7 T, \cite{aoki98}.\ At slightly higher field, H = 38 T yet
another large anomaly develops in C$_{m}$/T. This anomaly is suppressed with a
magnetic field of 40T, and its origin is unknown at the present time. Figure 1
(b) inset shows the C$_{m}$/T measured at H = 40 T, 42 T, and 45 T. In this
regime all that is left in C$_{m}$ is a Schottky-like anomaly that
shifts from T$_{max}\sim$5.6 K to 7.4 K ($\Delta$T$_{max}$ = 2K, $\sim$35$\%$)
when the magnetic field is increased by only 2 T (5$\%$). We fitted our data
with an expression for a Schottky anomaly using the following parameters:
$\Delta_{40T\text{ }}=$ 1.55 meV, $\Delta_{42T\text{ }}=$ 2.03 meV,
$\Delta_{45T\text{ }}=$ 2.48 meV, and degeneracy equal to 0.6, giving an
associated entropy $\sim$(0.3$\pm$ 0.02)$\times$R. Both T$_{max}$ and $\Delta$
point to possible singlet \textit{f}-electron crystal electric field (CEF) levels that cross at H
$\simeq$ 36-38 T. Such a level crossing was proposed as a semiquantitative
explanation for the observed phenomenology of URu$_{2}$Si$_{2}$ at high fields
\cite{santini98}. The upturn in C$_{m}$/T vs T at H > 42 T could be
due to a phonon component that differs from that of ThRu$_{2}$Si$_{2}$.

In order to follow the anomalies observed in these experiments down to mK
temperatures, we polished a bar of sample $\sharp$1 to dimensions 0.15 x 0.4 x
3mm$^{3}$ with its longer side along the crystallographic $c$-axis and
attached four gold leads using a spot welder, for magnetoresistance in pulsed
fields. The electrical contact resistance when prepared in this way resulted
in $\simeq$ 0.1 $\Omega$ each. We then mounted the sample on our Si/sapphire
sample holder parallel to the direction of the applied magnetic field, such as
to have H // $c$ // $i$, where $i$ is the applied electrical current. The
small mass and large area of the sample helps keeping the temperature constant
during pulses. For these measurements we used a capacitor driven pulsed magnet
able to produce a 400 ms pulse, and magnetic fields up to 50 T, at the
NHMFL-Los Alamos. The sample resistivity ($\rho$) was measured using a
standard lock-in amplifier detection technique operating at 173.2 kHz, and an
excitation current of not more than 4 mA. Figure 2 displays $\rho$ vs. H at
constant temperature for temperatures ranging from 0.5 K to 20K (bottom to
top). Curves were displaced for clarity. Only a broad maximum around 40 T is
observed above the HO phase transition T$_{o}\simeq$17K, possibly related to
the onset of metamagnetism \cite{sugiyama90}, but below T$_{o}$ a clear
minimum appears in $\rho$ vs. H. The minimum shifts monotonically to higher
fields as the temperature is reduced. We also observe a broad bump which
narrows at higher fields, until at 4.2K the resistance abruptly changes shape.
Here and below we start seeing three anomalies, first a drop, then an
increase, and finally a large drop in the sample resistance. Our higher
temperature data agrees partially with previous results \cite{devisser86}.
Note that while the magnitude of the resistivity obtained with the \emph{ac}
technique used in this study may be slightly affected by capacitive/inductive
effects, the magnetic fields at which anomalies are observed are not.

We have compiled all our data in Figure 3. In Figure 3(a) we have a phase
diagram for sample $\sharp$1 where we plotted the temperature and magnetic
fields at which we observe anomalies in C$_{m}$/T vs. T extracted from Fig.1
(solid symbols) and anomalies in $\rho$ vs. H curves extracted form Fig. 2
(open symbols). We establish in this figure the new high field phase in
URu$_{2}$Si$_{2}$ (region III), and note that the corresponding critical
fields extrapolated to zero temperature are H$_{o}$(0) = (35.9 $\pm$0.35) T
for the transition between regions (I) and (II), H$_{1}$(0) = (36.1$\pm$0.35)
T, for the transition between regions (II) and (III) and H$_{2}$(0) =
(39.7$\pm$0.35) T for the transition between regions (III) and (IV). Within
experimental error we find H$_{o}$(0) = H$_{1}$(0), a fact that could be
coincidental or, more interestingly, could indicate that regions (I) and (III)
are closely related. Our phase diagram shares some aspects with one proposed
before \cite{sakakibara93}, derived from susceptibility measurements at 1.3 K $\leq$ T $\leq$ 4.2 K.

In addition to the specific heat data, we measured the temperature changes in
URu$_{2}$Si$_{2}$ due to the magnetocaloric effect (MCE) during magnetic field
sweeps across the metamagnetic transition. Figure 3(a) Inset shows MCE\ data
taken at 3.5 K sweeping the magnetic field from 25 to 45 T, and then back to
25 T at a constant rate of $\simeq$12 T/min. Here we observe three
\emph{reversible} features, \emph{i.e.} they change sign with the field ramp.
When the magnetic field is increasing we see a temperature drop at H$_{o}$ =
34.5 T, then a peak at H$_{1}$ = 36.5 T\ and another drop at H$_{2}$ = 39 T.
\ We observe peaks, instead of steps, because of the calorimeter's finite
relaxation time constant $\tau_{cal}$. Since the total entropy of sample and
stage should be conserved within times t $< \tau_{cal}$, a temperature 
drop at H$_{o}$ implies an increase of magnetic
entropy (S$_{m}$) in the sample. The peak in the temperature at H$_{1}$
indicates a drop in S$_{m}$, and the second drop at H$_{2}$ another increase
in S$_{m}$. Our results suggest that the metamagnetic transition in URu$_{2}%
$Si$_{2}$ is accompanied by dramatic electronic band structure effects, in
which the density of available quantum states increases causing the greater
entropy of the system. Furthermore, the two features observed at higher fields
strongly indicate that we cross through region (III), described above. Figure
3 (b) exhibits the temperature changes observed in sample $\sharp$2, when
during the down-sweep, various initial temperatures were used. For this sample
we see the same features observed in sample $\sharp$1 at slightly different
fields, confirming that all results previously discussed are sample
independent. We notice kinks in the temperature vs. field that indicate the
high field and low field phases converge to the same critical field ($\sim$36
T) at zero temperature. We also see a small anomaly (connected by dashed line)
where we see the step-like feature in the C$_{m}$/T vs. T of sample $\sharp$1
displayed in Fig. 1(b). We note that while the jump in temperature at
H$_{2}\sim$39 T is sharp, suggesting a first order-like transition in field,
the anomalies at H$_{o}$ $\sim$35.5 T and H$_{1}\sim$36 T are more
rounded,\emph{\ i.e.} second order-like in field. A plot of temperature traces
during the field up-sweep has similar characteristics, without showing
significant field hysteresis.

To explain the observed properties we analyzed two different cases:
\textbf{A)}\ The normal state of URu$_{2}$Si$_{2}$ is the coherent heavy
fermion state in which \textit{f}-electrons acquire itinerant character upon
hybridization with conduction electrons. In this itinerant band scenario the
HO phase in region (I) is destroyed by a magnetic field due to Zeeman
splitting of spin-down and spin-up bands, and a new single-spin phase is
stabilized in region (III). Ordered states that may be affected in this way
are those that involve singlet pairing at a characteristic
translational\textbf{\ }(\textit{nesting}) wavevector \textbf{Q,} such us a
charge density wave, or the recently proposed incommensurate orbital
antiferromagnetic phase \cite{chandra02}. Region (III) may then be the
re-entrance of the HO phase with a different nesting wavevector, or an orbital
flopped phase. The Fermi surface instabilities produced by the Zeeman effect
may also explain the steps observed in the magnetization vs. field
\cite{deboer86} as new phases are stabilized. \textbf{B)}\ Localized
\textit{f}-electrons dominate the low temperature behavior of URu$_{2}$%
Si$_{2}$ and the phase transitions near 40 T are a consequence of crossing
singlet \textit{f}-electron crystal electric field (CEF) levels, as proposed
by Santini. Using a reduced quadrupolar coupling parameter $\lambda$
\cite{santini98}, it can be shown \cite{batista} that region (III) may be a
magnetic field induced antiferromagnetic quadrupolar phase. We observe two
problems with this scenario. First, region (I) remains unexplained. Second, a
Schottky contribution to the specific heat should be observed on both sides of
the level crossing field, which is not supported by our experiments. A
bicritical point\ in URu$_{2}$Si$_{2}$, a temperature below which a
coexistence line separates regions (I) and (III) in the phase diagram, may
exist below\textit{\ }0.5 K and thus quantum fluctuations (a quantum critical
point) may control the macroscopic thermodynamic and transport properties.
Further experiments near H$_{o}=$36 T are under way.

In summary, we determined systematically the low temperature/high magnetic
field phase diagram of URu$_{2}$Si$_{2}$ with measurements of specific heat vs
temperature in continuous magnetic fields up to 45 T, magnetocaloric effect
measurements, and magnetoresistance measurements in pulsed magnetic fields at
temperatures from 0.5 K to 20 K for the first time. The specific heat anomaly observed at the
onset of the HO phase T$_{o}$ $\simeq$ 17 K is completely suppressed in a
magnetic field of 36 T, and a new phase is revealed between 36 and 40 T in
which C$_{m}$/T shows a sharp first-order-like anomaly at T = 5.2 K. At 40-45 T 
no magnetic phase transition is observed, and C$_{m}$/T is dominated by a Schottky-like
contribution. We also show that the magnetocaloric effect can be used to study
the high field phases in detail.

We acknowledge N. Harrison, C. Batista, P. Chandra, P. Coleman, S. Crooker and
R. Movshovich for useful discussions and a critical reading of the manuscript,
A.A. Menovsky for providing sample $\sharp$1, D. Hinks, B. Sarma and A. Suslov
for providing sample $\sharp$2 and assistance during MCE experiments.
Experiments performed at the National High Magnetic Field Laboratory are
supported by the U.S. National Science Foundation through Cooperative Grant
No. DMR 9016241, the State of Florida and the U.S. Department of Energy.

\newpage%

%TCIMACRO{\TeXButton{Fig. 1}{\begin{figure}
%\caption{ {(a) C$_{m}$/T vs. T for magnetic fields up to 33.5 T in
%sample $\sharp$1. The solid line indicates large delta T method. Dotted
%lines are guides to the eye. (b) C$_{m}$/T vs. T for H = 36 T, no
%sharp anomaly present, and H = 38 T where a new anomaly is evident. Inset:
%C$_{m}$/T vs. T for H = 40, 42, 45 T. Dashed lines indicate fits with the
%Schotkky expression.}
%\label{ fig1}
%\end{figure}}}%
%BeginExpansion
\begin{figure}
\caption{(a) C$_{m}$/T vs. T for magnetic fields up to 33.5 T in
sample $\sharp$1. The solid line indicates large delta T method. Dotted
lines are guides to the eye. (b) C$_{m}$/T vs. T for H = 36 T, with no
sharp anomaly present, and H = 38 T where a new anomaly is evident. Inset:
C$_{m}$/T vs. T for H = 40, 42, 45 T. Dashed lines indicate fits with the
Schotkky expression. }
\label{ fig1}
\end{figure}%
%EndExpansion
%

%TCIMACRO{\TeXButton{Fig. 2}{\begin{figure}
%\caption{Magneto-resistivity of sample $\sharp$1 at constant temperature.
%All curves, except for T = 0.5 K, were displaced for clarity purposes.
%Anomalies associated to phase boundaries are indicated by arrows. }
%\label{ fig2}
%\end{figure}}}%
%BeginExpansion
\begin{figure}
\caption{Magneto-resistivity of sample $\sharp$1 at constant temperature.
All curves, except for T = 0.5 K, were displaced for clarity purposes.
Anomalies associated to phase boundaries are indicated by arrows. }
\label{ fig2}
\end{figure}%
%EndExpansion
%

%TCIMACRO{\TeXButton{Fig. 3}{\begin{figure}
%\caption{(a) Phase diagram for sample $\sharp$1 . ($\blacksquare$) specific
%heat maximum in the low fields regime, ($\bigstar$) step-like transition,
%($\blacktriangledown$) intermediate field peak, ($\bullet$) position of
%Schottky anomaly at higher fields. ($\square$, $\triangle$, $\bigcirc$)
%anomalies in the resistance vs. H. Inset: magnetocaloric effects sweeps. (b)
%Magnetocaloric effect in sample $\sharp$2. Darker shade indicates where
%transitions are sharper.}
%\label{fig3 }
%\end{figure}}}%
%BeginExpansion
\begin{figure}
\caption{ (a) Phase diagram for sample $\sharp$1 . ($\blacksquare$) specific
heat maximum in the low fields regime, ($\bigstar$) step-like transition,
($\blacktriangledown$) intermediate field peak, ($\bullet$) position of
Schottky anomaly at higher fields. ($\square$, $\triangle$, $\bigcirc$)
anomalies in the resistance vs. H. Inset: magnetocaloric effects sweeps. (b)
Magnetocaloric effect in sample $\sharp$2. Darker shade indicates where
transitions are sharper.}
\label{fig3 }
\end{figure}%
%EndExpansion

\end{document}